\documentclass{JHEP}

\newcommand{\be}{\begin{equation}}
\newcommand{\ee}{\end{equation}}

\newcommand{\bea}{\begin{eqnarray}}
\newcommand{\eea}{\end{eqnarray}}
\newcommand{\bean}{\begin{eqnarray*}}
\newcommand{\eean}{\end{eqnarray*}}

\def\beq{\begin{equation}}

\def\eeq{\end{equation}}

\def\tr{\mathop{\rm tr}}

\preprint{ {\tt hep-th/0303144}}

\title{Note on Seiberg Duality in Matrix Model}

\author{Bo Feng \\ Institute for Advanced Study \\
Einstein Drive, \\
Princeton, New Jersey, 08540 \\
email: fengb@sns.ias.edu
}

\abstract{In this note, we give a method to derive the Seiberg
duality by the matrix model. The key fact we used is that
the effective actions given by matrix model method should be
identical for both electric and magnetic theories. We demonstrate
our method for SQCD with $U(N)$, $SO(N)$ and $Sp(N)$ gauge groups.}
\keywords{Matrix Model, Seiberg Duality}
\begin{document}

\section{Introduction and Motivation}

The field theory {\sl v.s.} matrix model conjecture proposed 
by Diikgraaf and Vafa
\cite{Dijkgraaf:2002fc,Dijkgraaf:2002vw,Dijkgraaf:2002dh} has intrigued
a lot of works from various perspectives. The original idea comes from
the relationship between field theory and string theory, but later the
conjecture is proved by pure field theory methods in 
\cite{Dijkgraaf:2002xd,Cachazo:2002ry} for adjoint matter and
in \cite{Seiberg:2002jq} for massive fundamental flavors and adjoint matter
(The generalization to massless flavors has been given in \cite{Feng:2002is}
based on the work of Seiberg \cite{Seiberg:2002jq}). With these achievements,
matrix model becomes another alternative way to investigate many 
interesting problems in fields theories, like the new duality 
demonstrated in \cite{Cachazo:2002zk} (the generalization to
other cases in \cite{Ahn:2003cq,Brandhuber:2003va,Balasubramanian:2003tv})
and related works in \cite{Ferrari:2002kq,Ferrari:2003yr}.

Besides these successes of matrix model, we also like to know the limit
of the new method. The baryonic deformation has been addressed in
\cite{Argurio:2002hk,Bena:2002ua,Suzuki:2002jc,Bena:2003vk}
where it has been showed that although the baryonic deformation 
complexes the boundary condition in matrix model, there is a way to
sum up  relative contributions  for field theory in matrix model
 expansion. 
The multi-trace deformation was investigated in 
\cite{Balasubramanian:2002tm,Alishahiha:2003by} where it was pointed
out \cite{Balasubramanian:2002tm} that the direct matrix model 
integration of multi-trace deformation does not give back 
correct  results in field theory, but by linearization trick we
can reduce the multi-trace problem to the single-trace problem.
Except the adjoint and fundamental flavors, other matter contents
have been considered in \cite{Kraus:2003jf} where it was found that
the conjecture failed with these more general matter fields.
For example, the gauge theory $Sp(N)$ with antisymmetric chiral fields agrees 
with the matrix model up to $N/2$ loops in the perturbative theory,
but discrepancy shows at $N/2+1$ loops. We also like to ask what is
the correct matrix model description (if it exists) for chiral theories
because of their role in phenomenology.

The question we like to address in the note is the Seiberg duality
in matrix model. Seiberg duality of ${\cal N}=1$ theories 
\cite{Seiberg:1994pq,Intriligator:1995id,Intriligator:1995er,Intriligator:1995au}  is
a very nontrivial statement above two different UV theories in IR.
It states that these two theories (the electric
theory and the magnetic theory) will flow to same (nontrivial) conformal
fixed point in IR. With the new method of matrix model, it is natural to
apply to the Seiberg duality. In \cite{Feng:2002zb,Feng:2002yf},
explicit calculations in matrix models have been done for both 
electric and magnetic theories of SQCD with mass deformations of quarks
and it has been shown that the effective actions are same for
both theories, thus checked the Seiberg duality. Generalizations to
$SO/Sp$ groups are given in  \cite{Ahn:2002vj,Ahn:2003fq}.

However, as we emphasized in \cite{Feng:2002yf}, these calculations
serve as the check of Seiberg duality and we want to ask more
profound question: could we derive the Seiberg duality from the matrix
model? If we could, matrix model will be another powerful tool
to study the duality in field theory.

Let us analyze this question. The first idea to derive Seiberg
duality in matrix model is to try to find a proper transformation
of matrix superpotential in one theory (for example the electric
theory). However, it seems this naive method does not work.
There are several reasons. First, familiar transformations
(like the Legendre transformation) change one theory into
another equivalent theory while the dual pair are total
different UV theories. This can be seen from another point
of view. The dual pair will contribute to same effective
action in IR, while  the effective action in IR is not directly
related to the free energy of matrix models, but through
\be \label{W-F}
W_{eff}= N_c {\partial {\cal F}_{\chi=2}(S,g) \over \partial
S}+ {\cal F}_{\chi=1}(S,g)
\ee 
The relationship (\ref{W-F}) shows that if $W_e=W_g$, 
with general different $N_c$ for dual pair we will have
${\cal F}_e\neq {\cal F}_g$, i.e., they are two different matrix
theories with total different free energies.

The second reason can be also seen  from  (\ref{W-F}) that
the matrix model does not have any memory about the rank of
gauge group. We recover the information of rank only when
we go from the free energy to the effective action
where the rank $N_c$ appears as a multiplier. It tells us
that we should {\sl not} seek the transformation of Seiberg duality 
in matrix model at the level of free energy (or the 
matrix model superpotential), but at the level of effective
action. 
More concretely, starting with  two matrix models with superpotential
$W_{e,tree}$ and $W_{g,tree}$, we do the independent matrix
model integrations and calculate  effective actions
$W_{e,eff}$ and $W_{g,eff}$. These effective actions will be
functions of  glueball field $S$ and other fields as well as
 coupling constants. The idea is that if we 
require  $W_{e,eff}\equiv W_{g,eff}$
as  functions of all variables, we may derive the Seiberg duality.
We will show the idea works, at least for these examples we 
will discuss in this note.

%%%%%%%%%%%%%%%%%%%%%%%%%%%%
\section{The Seiberg dual theory of $U(N_c)$ group}

The theory we want to discuss is the $U(N_c)$ gauge group with
$N_f$ flavors $Q_i, \widetilde{Q}^i$\footnote{Various results in the SQCD like 
${\cal N}=1$ theory with $U(N)$ gauge group in matrix model
can be found in \cite{UN}.} and arbitrary deformation
$W_{tree}=V(M)$ of meson fields 
$M_i^j=Q_i^\alpha\widetilde{Q}^j_\alpha$ where $\alpha$ is color
index. The matrix model integration of the prototype has been
given in \cite{Demasure:2002sc} by the insertion of delta-function
 $\delta(M_i^j-Q_i^\alpha\widetilde{Q}^j_\alpha)$
with results as
\bea\label{generic}
W_{eff}(S,M) & = & (N_c-N_f) S[1-\log{ S\over \Lambda^3}]-S \log
\Bigl({ \det(M) \over \Lambda^{2N_f}}\Bigr)+V_{tree}(M)  
\eea
This is a pretty neat result because usually we can not do the matrix
model integration exactly\footnote{The matrix model integration 
of delta-function requires that the rank $M$ of matrix is larger than
the number $N_f$ of flavors. Since we have kept $N_f$ fixed while
taking the large $M$ limit in the matrix model integration, the
condition is satisfied.}. For this simple example with arbitrary
deformation of $V(M)$, (\ref{generic}) is exact. As a simple
exercise
we can take $V(M)=m_j^i Q_i^\alpha\widetilde{Q}^j_\alpha=\tr(m M)$
which has been done explicitly in  \cite{Feng:2002zb}. 
Equation (\ref{generic}) gives
\bean
W= (N_c-N_f) S[1-\log{ S\over \Lambda^3}]-S \log
\Bigl({ \det(M) \over \Lambda^{2N_f}}\Bigr)+\tr(m M)
\eean
Integrated out $M$ by
\bean
{\partial W\over \partial M}=0= -S M^{-1}+m
\eean
we get 
\be
W= N_c S[1-\log{ S\over \Lambda^3}]-S \log\Bigl( {\Lambda^{N_f} \over
\det(m)} \Bigr)
\ee
which matches the result in \cite{Feng:2002zb}.

Now we will apply above general result given by Demasure and Janik to
our Seiberg dual pair. Given the electric theory as above with
arbitrary deformation $V(M)$, we try to find the proper magnetic
theory $U(\widetilde{N}_c)$ with $N_f$ flavors $q_i,\widetilde{q}^i$,
singlets $M$ 
and proper superpotential $V(q,\widetilde{q},M)$. The first step we need
to do is to integrate the magnetic matrix model.  Here we have
fields $q_i,\widetilde{q}^i$ and gauge singlets $M$. Should we integrate
them all in matrix model? The answer is no. We need only integrate
fields  $q_i,\widetilde{q}^i$ in matrix model while keeping $M$ as
parameters. It is because fields $M$ are gauge singlets. So 
according to the field theory analysis
in \cite{Dijkgraaf:2002xd,Cachazo:2002ry,Seiberg:2002jq}, we should
leave $M$ untouched at the level of free energy and add them back to
the effective action directly by the prescription  (\ref{W-F}). 
This point has also  been  
emphasized in \cite{Balasubramanian:2002tm,Feng:2002is}. Using this
new understanding, we redo the integration of magnetic matrix
model in \cite{Feng:2002zb,Feng:2002yf} at Appendix to show the
consistence.

Since we do not need to integrate fields $M$, the matrix model
integration of magnetic theory is same prototype as discussed by
 Demasure and Janik and we can write down the effective superpotential
directly as
\be \label{generic-mag}
W_{g,eff}(S,M,\widetilde{M})=(\widetilde{N}_c-N_f) \widetilde{S}[1-\log{ \widetilde{S}
\over \widetilde{\Lambda}^3}]-\widetilde{S} \log
\Bigl({ \det(\widetilde{M}) \over \widetilde{\Lambda}^{2N_f}}\Bigr)
+V(M,\widetilde{M})
\ee
where to distinguish the magnetic theory from the electric theory, 
we use tilde for our
notations in 
magnetic theory and $\widetilde{M}_i^j$ is the magnetic meson given by
$q_i\cdot \widetilde{q}^j$. To compare with the electric theory
(\ref{generic}) we need to integrate out the magnetic meson 
$\widetilde{M}$.

Now it comes to the key point. Since we require
$W_{e,eff}=W_{g,eff}$ for {\sl arbitrary} deformation $V(M)$, it is
conceivable that we should have $V(M,\widetilde{M})=V(M)+f(M,\widetilde{M})$
where $f(M,\widetilde{M})$, which  describes the interaction of 
$M$ and $q_i\cdot \widetilde{q}^j$, does not depend on the deformation $V(M)$.
Because $M$ is gauge singlet and adjoint under the flavor symmetry
$U(N_f)$, the interaction of $M$ and 
$q_i\cdot \widetilde{q}^j$ should be like $\sum\tr(M^{p_1} \widetilde{M}^{q_1}
M^{p_2} \widetilde{M}^{q_2}...)$. Integrating out the magnetic meson 
$\widetilde{M}$, we have equation
\be \label{solve-tilde-M}
{\partial W_g \over \partial \widetilde{M}} =0 =
-\widetilde{S} \widetilde{M}^{-1}+{\partial f(M,\widetilde{M})\over
\partial\widetilde{M}}
\ee
From (\ref{solve-tilde-M}) we suppose to solve $\widetilde{M}$, put it
back to $W_{g,eff}$ and compare with  $W_{e,eff}$. Especially we should
have term $S \log(det(M))$ by putting  $\widetilde{M}$ back to
term  $\widetilde{S} \log(det(\widetilde{M}))$. It is hard to imagine
we can have this result unless the solution is 
$ \widetilde{M}^{-1} \sim M^{n}$. In another word, 
\be \label{assumption}
 f(M,\widetilde{M})= \tr(\widetilde{M} {M^n \over \mu^{2n-1}})
\ee
where $\mu$ is a scale constant. Under this assumption, we have
\be
\widetilde{M}^{-1}={ M^n \over \widetilde{S} \mu^{2n-1}}
\ee
Putting it back to $W_{g,eff}$ and simplifying, we get
\be \label{sim-mag}
W_{g,eff}=n \widetilde{S} \det(M)+\widetilde{N}_c \widetilde{S}
-\widetilde{N}_c \widetilde{S} \log  \widetilde{S}
+ \widetilde{S}\log { \widetilde{\Lambda}^{3\widetilde{N}_c-N_f}
\over   (\mu^{2n-1})^{N_f}}
\ee
where we have neglected the term $V(M)$ in $W_{g,eff}$ (we will
neglect the same term in  $W_{e,eff}$). The result should be
compared with the effective action of electric  theory
\be \label{sim-ele}
W_{e,eff}= -S \det(M)+ (N_c-N_f) S - (N_c-N_f) S \log S+ S \log
\Lambda^{3N_c-N_f}
\ee
which is just regrouped of equation (\ref{generic}). Comparing the
first term of (\ref{sim-mag}) and (\ref{sim-ele}) we get the first
condition
\be \label{map-1}
-S= n \widetilde{S}
\ee
Using (\ref{map-1}) to second and third terms we get
\be \label{map-2}
\widetilde{N}_c= n (N_f-N_c)
\ee
From this we see that $n$ must be positive integer. Comparing the
last term we get
\be \label{map-3}
\Lambda^{3N_c-N_f} (\widetilde{\Lambda}^{3\widetilde{N}_c-N_f})^{1\over n}
=(-n)^{-\widetilde{N}_c \over n} (\mu^{2n-1})^{N_f\over n}
\ee
Now it is clear that when $n=1$, equations (\ref{map-1}), (\ref{map-2})
and (\ref{map-3}) are exactly the dual relations of Seiberg dual pair.
Notice that just by requiring the matching of $W_{e,eff}$ and $W_{g,eff}$
we can not exclude the possibility $n\neq 1$. However, from 
(\ref{map-1}) we see that when $n \neq 1$, $|S|\neq |\widetilde{S}|$,
so it is very natural to choose $n=1$. In fact by the symmetry 
of dual pair and  the dual theory of the dual theory will go back to
original theory, we should choose $n=1$. To see this, notice that
$$
S\rightarrow [\widetilde{S}=-{S\over n}]\rightarrow
[\widetilde{\widetilde{S}}=-{\widetilde{S}\over n}= {S\over n^2}]~.
$$

\section{The Seiberg dual theory of $SO(N)$ and $Sp(N)$ groups}
 
The checking of Seiberg duality in matrix model for $SO(N)$ gauge group
with $N_f$ flavors $Q^j$\footnote{Other works of $SO/Sp$ groups in matrix
model can be found also in \cite{SOP}.}
 under the non-degenerated mass deformations
has been done in \cite{Ahn:2002vj}. The procedure to derive the 
Seiberg duality will be parallel to $U(N)$ case. Using the
delta-function technique, the general effective superpotential under
arbitrary meson deformation $V(M)$ with $M=Q^j\cdot Q^j$  is given by
 \cite{Ahn:2002vj}
\be \label{gen-ele-SO}
W_{e,eff}={1\over 2} (N_c-2-N_f) S[1-\log {S\over \Lambda^3}]
-{S\over 2} \log {\det(M) \over \Lambda^{2N_f}} +V(M) 
\ee
To see this, choosing $V(M)={1 \over 2} \tr(m M)$ and minimizing 
$W_{e,eff}$ in (\ref{gen-ele-SO}) regarding to $M$ we get
\bean
{\partial W_{e,eff} \over \partial M}= {-S\over 2} M^{-1}+ {m\over 2}=0
\eean
Putting it back to $W_{e,eff}$  and simplifying we get
\bean
W_{e,eff}={S\over 2} (N_c-2) [1-\log { S\over ( \Lambda^{3(N_c-2)-N_f}
\det(m)~~)^{1\over N_c-2}}]
\eean
which  is the result got in  \cite{Ahn:2002vj}. Using similar arguments
( i.e., (1) $M$ should not be integrated in matrix model; (2) 
the matching for arbitrary deformation $V(M)$ and the term 
$S\log \det(M)$) for
the magnetic theory we will have
\be \label{gen-mag-SO}
W_{g,eff}={1\over 2} (\widetilde{N}_c-2-N_f) \widetilde{S}
[1-\log {\widetilde{S}\over \widetilde{\Lambda}^3}]
-{\widetilde{S}\over 2} \log {\det(\widetilde{M}) \over \widetilde{\Lambda}^{2N_f}} +V(M) +{1\over 2 \mu^{2n-1}} \tr(M^n \widetilde{M})
\ee
Integrating out meson field $\widetilde{M}$ we have
\be
{\partial W_{g,eff} \over \partial \widetilde{M}}=
{\widetilde{S} \over 2} \widetilde{M}^{-1}+{M^n \over 2 \mu^{2n-1}}=0
\ee
Solving $\widetilde{M}$ and putting it back we simplify the
effective action as (notice that we neglected the term $V(M)$)
\be \label{sim-mag-SO}
W_{g,eff}={n \widetilde{S} \over 2}\log \det(M) +{\widetilde{S} \over 2}
(\widetilde{N}_c-2) (1-\log \widetilde{S})+{\widetilde{S} \over 2}
\log{\widetilde{\Lambda}^{3(\widetilde{N}_c-2)-N_f} \over
(\mu^{2n-1})^{N_f}}
\ee
which should be compared with 
\be \label{sim-ele-SO}
W_{e,eff}= -{S\over 2}\log \det(M)+{S\over 2}(N_c-N_f-2)(1-\log S)+{S\over 2}
\log \Lambda^{3(N_c-2)-N_f}
\ee
From the first three terms we get
\be \label{SO-map-1}
-S=n \widetilde{S},~~~~~~~\widetilde{N}_c-2=n(N_f-(N_c-2)~)
\ee
and from the last term we get
\be \label{SO-map-2}
\Lambda^{3(N_c-2)-N_f} ({\Lambda}^{3(\widetilde{N}_c-2)-N_f})^{1\over n}
=(-n)^{-{\widetilde{N}_c-2 \over n}} (\mu^{2n-1})^{N_f\over n}
\ee
Similar reason as in $U(N_c)$ case tells us to choose $n=1$. In this
case, equations (\ref{SO-map-1}) and (\ref{SO-map-2}) are exactly
the dual relations of Seiberg dual pair with $SO(N)$ gauge group.
Notice that to compare  (\ref{SO-map-2}) with the result in field
theory \cite{Intriligator:1995id}, we need to set
\be
\Lambda^{3(N_c-2)-N_f}_{matrix}=16 \Lambda^{3(N_c-2)-N_f}_{field}
\ee
as noticed in \cite{Ahn:2002vj}.

Comparing above calculation of $SO(N_c)$ with the one of $U(N_c)$,
we see they are same if we make the following replacement 
$N_c\rightarrow N_c-2$. When we discuss the gauge group $Sp(N)$ we
just need to use the replacement $N_c\rightarrow N_c+2$. With 
this replacement we will simply write down  results. Unlike the
$SO(N)$ case where the meson fields $M=Q^i\cdot Q^j$ are symmetric,
for $Sp(N)$ (the rank $r$ of $Sp(N)$ is $N/2$) the meson fields
$M= Q^i_a Q^j_b J^{ab}$ is antisymmetric \cite{Ahn:2003fq} where 
$J_{ab}=i\sigma_2 \otimes {\bf 1}_{r\times r}$. The effective superpotential
under general meson deformation is
\be \label{gen-ele-Sp}
W_{e,eff}={1\over 2} (N_c+2-N_f) S[1-\log {S\over \Lambda^3}]
-{S\over 2} \log {\det(M) \over \Lambda^{2N_f}} +V(M) 
\ee 
Similar reason constraints the effective superpotential for the
dual magnetic theory to be
\be \label{gen-mag-Sp}
W_{g,eff}={1\over 2} (\widetilde{N}_c+2-N_f) \widetilde{S}
[1-\log {\widetilde{S}\over \widetilde{\Lambda}^3}]
-{\widetilde{S}\over 2} \log {\det(\widetilde{M}) \over \widetilde{\Lambda}^{2N_f}} +V(M) +{1\over 2 \mu^{2n-1}} \tr(M^n \widetilde{M})
\ee
Integrated out $ \widetilde{S}$ from (\ref{gen-mag-Sp}) and
comparing with (\ref{gen-ele-Sp}),
we get following dual relations from matrix model
for $Sp(N)$ gauge group
\bea
-S=n \widetilde{S},~~~~~~~\widetilde{N}_c+2=n(N_f-(N_c+2)~) \\
\Lambda^{3(N_c+2)-N_f} ({\Lambda}^{3(\widetilde{N}_c+2)-N_f})^{1\over n}
=(-n)^{-{\widetilde{N}_c+2 \over n}} (\mu^{2n-1})^{N_f\over n}
\eea
The requirement of two time dualities going back to original theory
picks up $n=1$ solution.

These examples we discussed in this paper are simple and standard. It will
be interesting to generalize above method to other dual theories found in
field theory, for example, the one discussed by Kutasov and Schwimmer
in \cite{Kutasov:1994xq,Kutasov:1995ve}. Unlike these did in this
paper for which general effective actions are known by matrix model,
we do not know  results for generalized Seiberg dual theories at this moment.
But if we manage to do it by matrix model, it should be possible to
derive the dual theory by the matrix model method. 

\section*{Acknowledgements}
This research is supported under the NSF grant PHY-0070928.

\appendix

\section{Matrix integration in magnetic theory}
For the simplest magnetic theory with mass deformation
\be \label{mag-simplest}
W_g= \tr(mM)+{1\over \mu} q_i M^i_j \widetilde{q}_j
\ee
the matrix integration has been done in \cite{Feng:2002zb,Feng:2002yf},
where we integrated all fields $q,\widetilde{q}$ as
well as the gauge singlet fields $M$. However, from the  field theory
analysis in \cite{Dijkgraaf:2002xd,Cachazo:2002ry,Seiberg:2002jq}
as well as emphasized in  \cite{Balasubramanian:2002tm,Feng:2002is},
we should only integrate fields  $q,\widetilde{q}$  in matrix model and
leave  terms which are gauge invariant to the effective superpotential.
This method has been used to generalize the work of Seiberg  
\cite{Seiberg:2002jq} with massive flavors 
to the case of massless flavors in  \cite{Feng:2002is} where as a by-product,
the original proposal of insertion of delta-function with fundamental
flavors \cite{Demasure:2002sc} has been explained (see also \cite{Ohta:2002rd}
from another point of view about the delta-function). With these new
understanding, we should redo the matrix model integration for above
magnetic superpotential (\ref{mag-simplest}). It is similar to the 
example given in \cite{Feng:2002is} , but we include following calculations
for completeness which can also be considered as another example for
the justification of the delta-function.

Now let us do the calculation. The matrix model integration for
$q,\widetilde{q}$ can be found in \cite{Feng:2002zb} where meson
fields ${M_i^j\over \mu}$ have been treated as  mass parameters. The
result is
\be \label{W_g_middle}
W_{g,eff}= \widetilde{N}_c \Bigl(\widetilde{\Lambda}^{3\widetilde{N}_c-N_f}
\det({M\over \mu})\Bigr)^{1\over  \widetilde{N}_c} + \tr(m M)
\ee
where the first term comes after integrating out the glueball field 
$\widetilde{S}$ and the second term, from the original tree level
superpotential without matrix model integration. Next step is to minimize
meson fields $M$. From (\ref{W_g_middle}) we have
\be
{\partial W_{g,eff} \over \partial M}=0=\Bigl(\widetilde{\Lambda}^{3\widetilde{N}_c-N_f}
\det({M\over \mu})\Bigr)^{1\over  \widetilde{N}_c} M^{-1}+m
\ee 
which gives us
\be
\det(M)^{\widetilde{N}_c-N_f \over \widetilde{N}_c}
=(-)^{N_f} \Bigl({\widetilde{\Lambda}^{3\widetilde{N}_c-N_f} \over \mu^{N_f}}
\Bigr)^{N_f\over  \widetilde{N}_c} (\det (m))^{-1} 
\ee
Putting them back we get
\bean
W_{g,eff}& = & (\widetilde{N}_c-N_f) \Bigl({\widetilde{\Lambda}^{3\widetilde{N}_c-N_f} \over \mu^{N_f}}\Bigr)^{1\over  \widetilde{N}_c} 
\det(M)^{1 \over \widetilde{N}_c} \\
& = & (\widetilde{N}_c-N_f)  \Bigl({\widetilde{\Lambda}^{3\widetilde{N}_c-N_f} 
\over \det(-\mu m)}\Bigr)^{1\over  \widetilde{N}_c-N_f}
\eean
which is exactly the correct effective superpotential of magnetic theory.

\bibliographystyle{JHEP}

\begin{thebibliography}{99}

%\cite{Dijkgraaf:2002fc}
\bibitem{Dijkgraaf:2002fc}
R.~Dijkgraaf and C.~Vafa,
%``Matrix models, topological strings, and supersymmetric gauge theories,''
Nucl.\ Phys.\ B {\bf 644}, 3 (2002)
[arXiv:hep-th/0206255].
%%CITATION = HEP-TH 0206255;%%

%\cite{Dijkgraaf:2002vw}
\bibitem{Dijkgraaf:2002vw}
R.~Dijkgraaf and C.~Vafa,
%``On geometry and matrix models,''
Nucl.\ Phys.\ B {\bf 644}, 21 (2002)
[arXiv:hep-th/0207106].
%%CITATION = HEP-TH 0207106;%%


%\cite{Dijkgraaf:2002dh}
\bibitem{Dijkgraaf:2002dh}
R.~Dijkgraaf and C.~Vafa,
%``A perturbative window into non-perturbative physics,''
arXiv:hep-th/0208048.
%%CITATION = HEP-TH 0208048;%%


%\cite{Dijkgraaf:2002xd}
\bibitem{Dijkgraaf:2002xd}
R.~Dijkgraaf, M.~T.~Grisaru, C.~S.~Lam, C.~Vafa and D.~Zanon,
%``Perturbative Computation of Glueball Superpotentials,''
arXiv:hep-th/0211017.
%%CITATION = HEP-TH 0211017;%%


%\cite{Cachazo:2002ry}
\bibitem{Cachazo:2002ry}
F.~Cachazo, M.~R.~Douglas, N.~Seiberg and E.~Witten,
%``Chiral Rings and Anomalies in Supersymmetric Gauge Theory,''
arXiv:hep-th/0211170.
%%CITATION = HEP-TH 0211170;%%

%\cite{Seiberg:2002jq}
\bibitem{Seiberg:2002jq}
N.~Seiberg,
%``Adding Fundamental Matter to ``Chiral Rings and Anomalies in Supersymmetric Gauge Theory'',''
arXiv:hep-th/0212225.
%%CITATION = HEP-TH 0212225;%%


%\cite{Cachazo:2002zk}
\bibitem{Cachazo:2002zk}
F.~Cachazo, N.~Seiberg and E.~Witten,
%``Phases of N = 1 supersymmetric gauge theories and matrices,''
JHEP {\bf 0302}, 042 (2003)
[arXiv:hep-th/0301006].
%%CITATION = HEP-TH 0301006;%%

%\cite{Ahn:2003cq}
\bibitem{Ahn:2003cq}
C.~h.~Ahn and Y.~Ookouchi,
%``Phases of N = 1 supersymmetric SO / Sp gauge theories via matrix model,''
arXiv:hep-th/0302150.
%%CITATION = HEP-TH 0302150;%%

%\cite{Brandhuber:2003va}
\bibitem{Brandhuber:2003va}
A.~Brandhuber, H.~Ita, H.~Nieder, Y.~Oz and C.~Romelsberger,
%``Chiral rings, superpotentials and the vacuum structure of N = 1  supersymmetric gauge theories,''
arXiv:hep-th/0303001.
%%CITATION = HEP-TH 0303001;%%

%\cite{Balasubramanian:2003tv}
\bibitem{Balasubramanian:2003tv}
V.~Balasubramanian, B.~Feng, M.~x.~Huang and A.~Naqvi,
%``Phases of N = 1 supersymmetric gauge theories with flavors,''
arXiv:hep-th/0303065.
%%CITATION = HEP-TH 0303065;%%

%\cite{Ferrari:2002kq}
\bibitem{Ferrari:2002kq}
F.~Ferrari,
%``Quantum parameter space and double scaling limits in N = 1 super  Yang-Mills theory,''
arXiv:hep-th/0211069.
%%CITATION = HEP-TH 0211069;%%

%\cite{Ferrari:2003yr}
\bibitem{Ferrari:2003yr}
F.~Ferrari,
%``Quantum parameter space in super Yang-Mills. II,''
arXiv:hep-th/0301157.
%%CITATION = HEP-TH 0301157;%%


%\cite{Demasure:2002sc}
\bibitem{Demasure:2002sc}
Y.~Demasure and R.~A.~Janik,
%``Effective matter superpotentials from Wishart random matrices,''
arXiv:hep-th/0211082.
%%CITATION = HEP-TH 0211082;%%

%\cite{Feng:2002zb}
\bibitem{Feng:2002zb}
B.~Feng,
%``Seiberg Duality in Matrix Model,''
arXiv:hep-th/0211202.
%%CITATION = HEP-TH 0211202;%%


%\cite{Feng:2002yf}
\bibitem{Feng:2002yf}
B.~Feng and Y.~H.~He,
%``Seiberg Duality in Matrix Models II,''
arXiv:hep-th/0211234.
%%CITATION = HEP-TH 0211234;%%

%\cite{Balasubramanian:2002tm}
\bibitem{Balasubramanian:2002tm}
V.~Balasubramanian, J.~d.~Boer, B.~Feng, Y.~H.~He, M.~x.~Huang, V.~Jejjala and A.~Naqvi,
%``Multi-Trace Superpotentials vs. Matrix Models,''
arXiv:hep-th/0212082.
%%CITATION = HEP-TH 0212082;%%

%\cite{Feng:2002is}
\bibitem{Feng:2002is}
B.~Feng,
%``Note on Matrix Model with Massless Flavors,''
arXiv:hep-th/0212274.
%%CITATION = HEP-TH 0212274;%%

%\cite{Ohta:2002rd}
\bibitem{Ohta:2002rd}
K.~Ohta,
%``Exact Mesonic Vacua From Matrix Models,''
arXiv:hep-th/0212025.
%%CITATION = HEP-TH 0212025;%%

%\cite{Argurio:2002hk}
\bibitem{Argurio:2002hk}
R.~Argurio, V.~L.~Campos, G.~Ferretti and R.~Heise,
%``Baryonic Corrections to Superpotentials from Perturbation Theory,''
arXiv:hep-th/0211249.
%%CITATION = HEP-TH 0211249;%%

%\cite{Bena:2002ua}
\bibitem{Bena:2002ua}
I.~Bena, R.~Roiban and R.~Tatar,
%``Baryons, Boundaries and Matrix Models,''
arXiv:hep-th/0211271.
%%CITATION = HEP-TH 0211271;%%

%\cite{Suzuki:2002jc}
\bibitem{Suzuki:2002jc}
H.~Suzuki,
%``Mean-field Approach to the Derivation of Baryon Superpotential from Matrix Model,''
arXiv:hep-th/0212121.
%%CITATION = HEP-TH 0212121;%%

%\cite{Bena:2003vk}
\bibitem{Bena:2003vk}
I.~Bena, H.~Murayama, R.~Roiban and R.~Tatar,
%``Matrix Model Description of Baryonic Deformations,''
arXiv:hep-th/0303115.
%%CITATION = HEP-TH 0303115;%%


%\cite{Kraus:2003jf}
\bibitem{Kraus:2003jf}
P.~Kraus and M.~Shigemori,
%``On the matter of the Dijkgraaf-Vafa conjecture,''
arXiv:hep-th/0303104.
%%CITATION = HEP-TH 0303104;%%

%\cite{Seiberg:1994pq}
\bibitem{Seiberg:1994pq}
N.~Seiberg,
%``Electric - magnetic duality in supersymmetric nonAbelian gauge theories,''
Nucl.\ Phys.\ B {\bf 435}, 129 (1995)
[arXiv:hep-th/9411149].
%%CITATION = HEP-TH 9411149;%%

%\cite{Intriligator:1995id}
\bibitem{Intriligator:1995id}
K.~A.~Intriligator and N.~Seiberg,
%``Duality, monopoles, dyons, confinement and oblique confinement in supersymmetric SO(N(c)) gauge theories,''
Nucl.\ Phys.\ B {\bf 444}, 125 (1995)
[arXiv:hep-th/9503179].
%%CITATION = HEP-TH 9503179;%%

%\cite{Intriligator:1995er}
\bibitem{Intriligator:1995er}
K.~A.~Intriligator and N.~Seiberg,
%``Phases of N = 1 supersymmetric gauge theories and electric-magnetic  triality,''
arXiv:hep-th/9506084.
%%CITATION = HEP-TH 9506084;%%

%\cite{Intriligator:1995au}
\bibitem{Intriligator:1995au}
K.~A.~Intriligator and N.~Seiberg,
%``Lectures on supersymmetric gauge theories and electric-magnetic  duality,''
Nucl.\ Phys.\ Proc.\ Suppl.\  {\bf 45BC}, 1 (1996)
[arXiv:hep-th/9509066].
%%CITATION = HEP-TH 9509066;%%


%\cite{Ahn:2002vj}
\bibitem{Ahn:2002vj}
C.~Ahn and S.~Nam,
%``Supersymmetric SO(N) Gauge Theory and Matrix Model,''
arXiv:hep-th/0212231.
%%CITATION = HEP-TH 0212231;%%

%\cite{Ahn:2003fq}
\bibitem{Ahn:2003fq}
C.~Ahn,
%``Supersymmetric SO(N)/Sp(N) Gauge Theory from Matrix Model:Exact Mesonic Vacua,''
arXiv:hep-th/0301011.
%%CITATION = HEP-TH 0301011;%%

%\cite{Alishahiha:2003by}
\bibitem{Alishahiha:2003by}
M.~Alishahiha and H.~Yavartanoo,
%``On the multi trace superpotential and corresponding matrix model,''
arXiv:hep-th/0303074.
%%CITATION = HEP-TH 0303074;%%

%\cite{Kutasov:1995ve}
\bibitem{Kutasov:1995ve}
D.~Kutasov,
%``A Comment on duality in N=1 supersymmetric nonAbelian gauge theories,''
Phys.\ Lett.\ B {\bf 351}, 230 (1995)
[arXiv:hep-th/9503086].
%%CITATION = HEP-TH 9503086;%%

%\cite{Kutasov:1994xq}
\bibitem{Kutasov:1994xq}
D.~Kutasov and A.~Schwimmer,
%``Universality in two-dimensional gauge theory,''
Nucl.\ Phys.\ B {\bf 442}, 447 (1995)
[arXiv:hep-th/9501024].
%%CITATION = HEP-TH 9501024;%%

\bibitem{UN}
N.~Dorey, T.~J.~Hollowood, S.~P.~Kumar and A.~Sinkovics, hep-th/0209099,
hep-th/0209089; N.~Dorey, T.~J.~Hollowood and S.~P.~Kumar, hep-th/0210239;
F.~Ferrari, hep-th/0210135; R.~Dijkgraaf, S.~Gukov, V.~A.~Kazakov and C.~Vafa,
hep-th/0210238; R.~Argurio, V.~L.~Campos, G.~Ferretti and R.~Heise, 
hep-th/0210291; J.~McGreevy, hep-th/0211009; H.~Suzuki, hep-th/0211052;
I.~Bena and R.~Roiban, hep-th/0211075; R.~Gopakumar, hep-th/0211100;
S.~Naculich, H.~Schnitzer and N.~Wyllard, hep-th/0211123, hep-th/0211254;
R.~Dijkgraaf, A.~Neitzke and C.~Vafa, hep-th/0211194;
H.~Itoyama and A.~Morozov, hep-th/0211259, hep-th/0212032; 
T.~J.~Hollowood, hep-th/0212065; S.~Seki, hep-th/0212079;
I.~Bena, S.~d.~Haro and R.~Roiban, hep-th/0212083; 
C.~Hofman, hep-th/0212095; Y.~Demasure and R.~A.~Janik, hep-th/0212212;
T.~Mansson, hep-th/0302077; C.~I.~Lazaroiu, hep-th/0303008;
D.~Berenstein, hep-th/0210183, hep-th/0303033.


\bibitem{SOP}
H.~Fuji and Y.~Ookouchi, hep-th/0210148, hep-th/0205301;
H.~Ita, H.~Nieder and Y.~Oz, hep-th/0211261; 
Y.~Ookouchi, hep-th/0211287; R.~A.~Janik and N.~A.~Obers, hep-th/0212069;
S.~K.~Ashok, R.~Corrado, N.~Halmagyi, K.~D.~Kennaway and C.~Romelsberger,
hep-th/0211291; B.~Feng, hep-th/0212010; C.~h.~Ahn and S.~k.~Nam,
hep-th/0301203; Y.~Ookouchi and Y.~Watabiki, hep-th/0301226;
R.~Abbaspur, A.~Imaanpur and S.~Parvizi, hep-th/0302083;
A.~Klemm, K.~Landsteiner, C.~I.~Lazaroiu and I.~Runkel, hep-th/0303032;
\end{thebibliography}

\end{document}